\documentstyle[12pt]{article}
\begin{document}
\title{INTERFERENCE OF CHARGED PARTICLES IN A VECTOR POTENTIAL
                    WITH VANISHING MAGNETIC FIELD}
\author{I.H.Duru\\
Trakya University,   
Mathematics Department
  Edirne, 
  Turkey \\
 and \\       
TUBITAK\thanks{Mailing address.
e-mail: duru@mam.gov.tr. }, 
 Research Institute for Basics Sciences, \\
 PO Box 21, 41470
Gebze, Turkey}
\date{April 1997}

\maketitle
\begin{abstract}
An interference experiment in a magnetic field free 
region with non vanishing
vector potential created by two perpendicularly intersecting 
planes carrying uniform
currents is discussed. The relation of this configuration 
to the Aharonov-Bohm potential
is examined. An experimental set up which is finite in the 
direction of the electronic
motion is studied.

\end{abstract}


\vspace{2cm}

\noindent
RIBS-PH-97/7

\noindent
hep-ph/9704433

\pagebreak
 
\section{INTRODUCTION}

According to quantum physics charged particles interact 
with the external
electromagnetic potentials even in the regions where the field 
strengths are zero. The
well known example is the Aharonov-Bohm (AB) effect 
where the electrons are
observed to scatter in the potential of confined magnetic flux
\cite{1,2,3}.
In another example
the roles of the flux line and the electrons are intercharged: Neutrons, 
which have
magnetic dipole moments, are scattered in the Coulomb field of the 
point charge \cite{4}.
This effect too, is experimentally verified \cite{5}. The third example 
one can mention is the
electric AB effect in which a homogeneous but time 
dependent electric potential acts
on the electrons \cite{2}. The electric AB effect is yet to be 
experimentally observed. One
can think of some other geometrical configurations which may 
give rise to AB type
effects.

Here we present a new example for a vector potential with zero magnetic 
field.
We notice that two perpendicular planes carrying uniform currents in 
the direction
perperdicular to the intersection create a region in which we have such 
a potential.
Although it looks quite different, the potential we propose is 
related to the original AB
potential of the magnetic flux line by a simple map: If we, 
represent our vector potential
$A$ and the AB potential $A_{\rm AB}$ as complex numbers they satisfy 
the relation $A A_{\rm AB}$ = 1.
If the electrons moving in the region of space having this potential 
are allowed to
interfere with the free electrons, hyperbolic pattern of maxima and 
minima are shown
to be created. We hope that it may be possible to realise an experimental 
setup for this
potential and then the scattering of electrons in it can be observed. 
This hope is the
main motivation for writing up the present note. In fact a brief 
discussion presented in
the final section shows that in a less idealized experimental 
system which is finite in the
direction of the electronic motion the effect should still 
manifest itself.

\section{REALIZATION of a NEW VECTOR POTENTIAL with VANISHING MAGNETIC 
    FIELD}

Consider uniform current densities $I$ on the $xz$ and $yz$--planes 
which flow in $-x$ and
$-y$ directions. The corresponding spatial current density $j(x)$ is           
\begin{equation}
\vec{j}=-I\left( \delta (y)\hat{x}+\delta (x)\hat{y}\right) .
\end{equation}
The resulting vector potential and the magnetic field (with $c=1$) are 
given by
\begin{equation}
\vec{A}=4\pi I\left( \theta (y)y\hat{x}+\theta (x)x\hat{y}\right) 
\end{equation}
and
\begin{equation}
\vec{H}=4\pi I\left( \theta (x)-\theta (y)\right) \hat{z} .
\end{equation}

Four spatial sectors are distinguished: In the regions defined 
by $x,\ y>0$
and  $x,\ y<0$
the magnetic field is zero. In the other regions defined
by $x>0,\ y<0$
and $x<0,\ y>0$
the magnetic field values are constant and given by
$\vec{H}=4\pi I \hat{z} $
and $\vec{H}=-4\pi I \hat{z} $
respectively.

The interesting region is the quarter of space defined by the 
positive $x$ and $y$--axis
where the magnetic field vanishes, while the potential is finite:
\begin{equation}
\vec{H}=0,\ \vec{A}=4\pi I(y,x,0);\ x,y>0 .
\end{equation}

The above potential is connected to the Aharonov-Bohm (AB) 
potential by a simple map:

The AB potential produced by the magnetic flux confined to 
the $z$--axis is given
by
\begin{equation}
\vec{A}_{AB}=\frac{\Phi }{2\pi} \left( 
\frac{y}{x^2+y^2}, \frac{x}{x^2+y^2},0
\right) .
\end{equation}
Let us represent our potential of (4) and the AB potential as 
the complex numbers as
\begin{equation}
A=4\pi I (x+iy);\  x,y>0
\end{equation}
and
\begin{equation}
A_{AB}=\frac{\Phi }{2\pi} 
\frac{1}{x^2+y^2} (x+iy)
\end{equation}
respectively. If we choose the value of the flux as
$\Phi =8\pi^2I,$
we observe that the above
complex numbers are related to each other by
\begin{equation}
A=\frac{1}{\bar{A}_{AB}}
\end{equation}
where
$\bar{A}_{AB}$
is the conjugate of (7). This relation is the same as the 
conformal map which
transforms inside of the unit cylinder to the outside region 
which however in our case
valid only for the quarter of space defined by the positive 
$x,y$--axis.

\section{AN IDEAL INTERFERENCE EXPERIMENT}

The Schr\"{o}dinger equation for a particle with mass $\mu$ 
and charge $e$ in the
region  $x,y>0$ is (with $\hbar =1)$ 
\begin{equation}
-\frac{1}{2\mu}\left[(\partial_x - 4i\pi eI)^2+(\partial_y -4i\pi
eIx)^2
+\partial_z^2\right] \Psi_{++}=i\frac{\partial }{\partial t}\Psi_{++} .
\end{equation}

The solution differs from the free wave function
$\Psi_0$ by a pure phase:
\begin{equation}
\Psi_{++}(\vec{x},t) =e^{4i\pi Iex_+y_+} \Psi_{0}(\vec{x},t)
\end{equation}
where $x_+,\ y_+$ stand for the coordinates in
$x,\ y>0$ region. 

Suppose a coherent electron beam which is prepared to move 
in the positive $z$--direction 
is split into two parts, and then are let to enter
the $x,\ y>0$
and $x,\ y<0$
regions [Fig.1]. The wave function of the beam moving
in $x,\ y>0$
region is given by
\begin{equation}
\Psi_{++} =e^{4i\pi Iex_+y_+} \Psi_{0}^k
\end{equation}
where
\begin{equation}
\Psi_0^k=\frac{1}{\sqrt{2\pi}}e^{-i\frac{k^2}{2\mu}t} e^{ikz}
\end{equation}
is the free wave function for the motion parallel to $z$--axis. 
Note that the phase picked
up by the electron beam remains constant, i.e., depends only 
on the transverse position
in the $xy$--plane, 
independent of the distance traveled in $z$--direction. 
The second beam
moving in $x,\ y<0$
region will simply be described by the free wave function:
\begin{equation}
\Psi_{--}=\Psi_0^k.
\end{equation}
     
If we recombine the above beams to interfere somewhere at the 
asymptotic
values of $z$--axis, the resulting wave function will be
\begin{equation}
\Psi =\frac{1}{\sqrt{2}}(1+e^{4i\pi Iex_+y_+})\Psi_0^k.
\end{equation}
The corresponding probability density is
\begin{equation}
\left| \Psi \right|^2=\left( 1+\cos (4i\pi Iex_+y_+)\right) \left|
\Psi_0^k\right|^2
\end{equation}
which is dependent on the position of the first beam in
$x,\ y>0$ region.

If we let the second beam coming from the free region
$x,\ y<0$ interfere 
with the first beam which traveled in the
$x,\ y>0$ region at a position with
coordinates $x_+$ and $y_+$ given by
\begin{equation}
x_+y_+=\frac{2n}{4eI};\  n=1,2,3, \dots
\end{equation}
we should observe a maximum.

On the other hand if the free beam interferes with a beam from
the $x,\ y>0$
region which traveled at a position
\begin{equation}
x_+y_+=\frac{2n+1}{4eI};\  n=1,2,3, \dots
\end{equation}
we must have a minimum.

Repeating the experiment with different transverse positions 
of the beam coming
from the $x,\ y>0$
region we should be able to observe the hyperbolic curves of the
interference pattern [Fig.2]
\begin{equation}
y_+=\frac{n/2eI}{x_+}.
\end{equation}

\section{DISCUSSIONS}

So far we have discussed an ideal arrangement involving 
infinite planes. Of
course, in practice``infinite" means the employment of distances 
which are sufficiently
large in comparison to the wave lengths of the particles. Since 
the phase in the wave
function (11) is independent of the distance traveled in the 
$z$--direction, we should still
expect to observe an interference effect for a set-up which is 
of finite size along the $z$--
direction.

Let us now consider such a system in greater detail. 
If the length of the system
$Z$ in this direction is large compared to the wave length
$\lambda =k_z^{-1}$
of 
the electron, for
example if
$Z>10^2\lambda$
we expect that in the central sections around
$z\sim Z/2$ the
potentials to be roughly the same as those in the ideal case. 
Then, the electron moving
in the transverse position
$x,\ y>0$
picks up the phase
\begin{equation}
e^{4i\pi Iex_+y_+}
\end{equation}
whereas the other branch of the electron beam travelling 
along the $z$--axis in any part of
the $x,\ y<0$ region acquires no phase in this central section.

If coherence between two paths of the electron beam is 
disturbed at the entrance
and the exit sections corresponding to
$z\sim 0$
and
$z\sim Z$
respectively the 
ideal wave
functions (11) and (13) are replaced by
\begin{equation}
\Psi_{++} =e^{4i\pi Iex_+y_+} \Psi_{0}^+
\end{equation}
and 
\begin{equation}
\Psi_{--}=\Psi_0^-.
\end{equation}
Here
$\Psi_0^+$
and
$\Psi_0^-$
are some functions different from
$\Psi_0^k$
of (12) due to the 
disturbances
caused by entrance and exit effects. 
If the proportion of
$\Psi_0^+$
and
$\Psi_0^-$
is given by a
function
$\beta (\vec{x})$
in the form
\begin{equation}
\Psi_0^+=\beta \Psi_0^-
\end{equation}
the probability density (15) must then be replaced by
\begin{equation}
\left| \Psi \right|^2=\frac{1}{2}
\left( 1+|\beta |^2+\beta e^{4i\pi Iex_+y_+}
+\beta^* e^{-4i\pi Iex_+y_+}\right) 
\left| \Psi_0^-\right|^2.
\end{equation}
We now consider two cases:

\noindent
(i) For real $\beta$ it is obvious that we have
\begin{equation}
\left| \Psi \right|^2=\left(
\frac{1+\beta^2}{2}
+\beta \cos (4i\pi Iex_+y_+)
\right) 
\left| \Psi_0^-\right|^2;\ \beta ={\rm real} ,
\end{equation}
thus, the hyperbolic interference pattern of Fig.2 is still 
valid provided that the sign
of $\beta$
does not depend on the transverse coordinates
$x_+,\ y_+.$

\noindent
(ii) If $\beta$ is a complex function given by
\begin{equation}
\beta =|\beta |e^{i\alpha}
\end{equation}
the probability density becomes
\begin{equation}
\left| \Psi \right|^2=\left(
\frac{1+|\beta|^2}{2}
+|\beta | \cos (4i\pi Iex_+y_+)
\right) 
\left| \Psi_0^-\right|^2 ,
\end{equation}
which for
$|\beta |$ and
$\alpha $ being slowly varying functions of the transverse
coordinates
$x_+,\ y_+$
we again have the same hyperbolic interference 
pattern modulated
by $\alpha .$

\begin{center}
{\bf ACKNOWLEDGMENT}
\end{center}

Author likes to thank A.N. Aliev, \"{O}.F. Dayi, Y. Nutku 
and C. Sa\c{c}l{\i}o\u{g}lu for
illuminating discussions. He is also grateful to 
F. Haciev for his suggestions on the
experimental aspects.

\end{document}